\RequirePackage{fixltx2e}
\documentclass[twocolumn, aps, prl, superscriptaddress, floatfix]{revtex4-1}

\usepackage[colorlinks]{hyperref}
\usepackage{lmodern}
\usepackage{amsfonts, amssymb, amsmath}
\usepackage{graphicx}
\usepackage{color}
\usepackage[usenames,dvipsnames]{xcolor}
\usepackage[utf8]{inputenc}

\definecolor{ElectricPink}{RGB}{248,38,244}

\newcommand{\br}{\mathbf{r}}
\newcommand{\bu}{\mathbf{u}}
\newcommand{\bx}{\mathbf{x}}
\newcommand{\bB}{\mathbf{B}}

\newcommand{\boeta}{\hbox{\boldmath $\eta$}}

\begin{document}
	
	\title{Inertial-Range Reconnection in Magnetohydrodynamic Turbulence\\ and in the Solar Wind}

	\author{Cristian C. Lalescu}
	\affiliation{Department of Applied Mathematics \& Statistics, The Johns Hopkins University, Baltimore, MD 21218, USA}
	
	\author{Yi-Kang Shi}
	\affiliation{Department of Applied Mathematics \& Statistics, The Johns Hopkins University, Baltimore, MD 21218, USA}
	
	\author{Gregory L. Eyink}
	\email{eyink@jhu.edu}
	\affiliation{Department of Applied Mathematics \& Statistics, The Johns Hopkins University, Baltimore, MD 21218, USA}
	\affiliation{Department of Physics \& Astronomy, The Johns Hopkins University, Baltimore, MD 21218, USA}
	
	\author{Theodore D. Drivas}
	\affiliation{Department of Applied Mathematics \& Statistics, The Johns Hopkins University, Baltimore, MD 21218, USA}
	
	\author{Ethan T. Vishniac} 
	\affiliation{Department of Physics \& Engineering Physics, University of Saskatchewan, Saskatoon, Saskatchewan S7N 5E2, Canada}
	
	\author{Alexander Lazarian}
	\affiliation{Department of Astronomy, University of Wisconsin, 475 North Charter Street, Madison, Wisconsin 53706, USA}  

	\begin{abstract}
	{\it In situ} spacecraft data on the solar wind show events identified as magnetic reconnection with outflows and apparent
	``$X$-lines'' $10^{3-4}$ times ion scales. To understand the role of turbulence at these scales, we make a case 
	study of an inertial-range reconnection event in a magnetohydrodynamic (MHD) simulation. We observe stochastic 
	wandering of field-lines in space, breakdown of standard magnetic flux-freezing due to Richardson dispersion, and 
	a broadened reconnection zone containing  many current sheets. The coarse-grain magnetic geometry is like large-scale 
	reconnection in the solar wind, however, with a hyperbolic flux-tube or ``X-line'' extending over integral length-scales. 
        \end{abstract}
	
	\pacs{140.27 Ak}

\maketitle

Magnetic reconnection is widely theorized to be the source of explosive energy release 
in diverse astrophysical systems, including solar flares and coronal mass ejections \cite{PriestForbes07}, 
gamma-ray bursts \cite{ZhangYan11}, and magnetar giant flares \cite{Lyutikov06}. Because of the 
large length-scales involved and consequent high Reynolds numbers, many of these phenomena are expected to occur 
in a turbulent environment, which profoundly alters the nature of reconnection \cite{LazarianVishniac99,Eyinketal11,
Eyinketal13}.  In the solar wind near 1 AU, which is the best-studied turbulent plasma in nature, quasi-stationary 
reconnection has been observed for magnetic structures at a wide range of scales, from micro-reconnection events 
at the scale of the ion gyroradius ($\sim$100 km), up to integral length scales 
($\sim 10^{4-5}$ km), and even to larger scales \cite{Gosling12}. Yet numerical 
studies of reconnection in magnetohydrodynamic (MHD) turbulence simulations have focused almost 
exclusively on small-scale reconnection at the resistive scale \cite{Servidioetal10, Zhdankinetal13, 
Osmanetal14}.
Our objective in this Letter is to identify an inertial-range reconnection event in an MHD turbulence simulation
and to determine its characteristic signatures, for comparison with observations in 
the solar wind and other turbulent astrophysical environments. 

To search for reconnection at inertial-range scales we adapt standard observational criteria employed for the 
solar wind. In pioneering studies, Gosling \cite{Gosling12} has looked for simultaneous large increments of magnetic field $\delta \bB(\br)$ 
and of velocity field $\delta \bu(\br)$ across space-separations $r$ near the proton gyroradius $\rho_p$, which 
approximate MHD rotational discontinuities. Candidate reconnection events are  then identified as pairs of such 
near-discontinuities, with $\delta \bB(\br)$ aligned for the two members of the pair and $\delta \bu(\br)$ anti-aligned. 
Gosling's selected events generally have the appearance of two back-to-back shocks, or a ``bifurcated current sheet''. 
We modify this criterion to allow for more gradual field-reversals, by choosing instead $r=L/10,$ with $L$ the 
outer (integral) length of the turbulent inertial range
and by considering pairs separated  
by distances up to $L/2.$ 

We apply the above 
criterion to two datasets. 
The first is from a numerical simulation of incompressible,
resistive MHD in a $[-\pi,\pi]^3$ periodic cube, in a state of stationary turbulence driven by a large-scale body-force. The simulation 
has about a decade of power-law inertial-range and the full output for a large-scale eddy turnover time is archived in an 
online, web-accessible database 
\cite{JHUTDB}.
The second dataset consists of {\it Wind} spacecraft observations of the solar 
wind magnetic field $\bB$, velocity $\bu$, and proton number density $n_p$.
The results presented here are from a week-long fast stream in days 14-21 of 2008
(cf. \cite{Gosling07}). The average 
solar wind conditions were $u=638$ km/s,  $B=4.3$ nT, $n_p= 2.3$ ${\rm cm}^{-3}$,  Alfv\'en speed $V_A =62$ km/s, 
proton beta $\beta_p=1.1,$ and proton gyroradius $\rho_p=154$ km. The temporal data-stream from the spacecraft
is converted to an equivalent space series using Taylor's hypothesis, $x= u t$ \cite{Taylor38}. Simulated ``spacecraft 
observations'' from the MHD database are taken along 192 linear cuts, with $4^3$ cuts through each face of the simulation cube in the 
three coordinate directions.  We find a good correspondence for statistics of $\delta \bB(\br)$ and $\delta \bu(\br)$ in the two datasets,
with the grid spacing $dx=2\pi/1024$ of the simulation related to $18$ s of the {\it Wind} time-series \cite{Supplement}. 
We thus estimate the turbulent outer scale $L$ of the solar wind stream 
to be 6.5$\times 10^5$ km, or $\sim$1026 s in time units, compared with $L=0.35$ in the MHD simulation. 

\begin{figure}[h!]
\includegraphics[width=0.8\columnwidth,height=0.7\columnwidth]{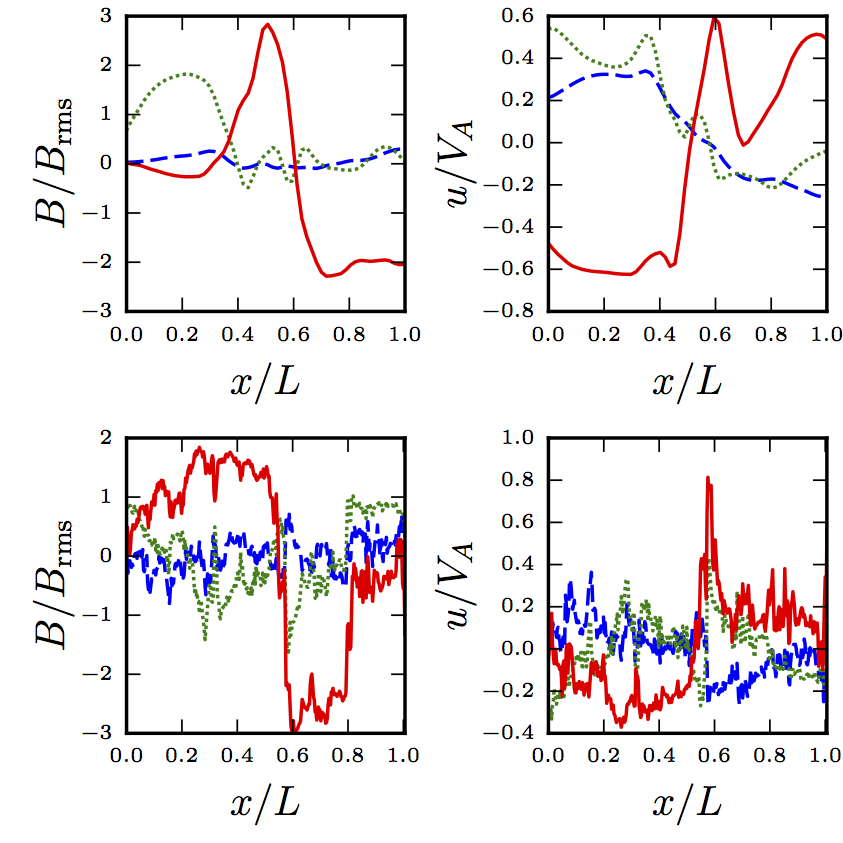}
\caption{{\it Top:} Event from the MHD simulation, at point (2.964,0.908,5.841) along a cut in 
the $y$-direction. Magnetic fields (left) normalized by $B_{rms} = 0.24,$ and velocity fields (right) 
by local upstream Alfven velocity $V_A= 0.7$. {\it Bottom}: Event from {\it Wind} spacecraft data, 
on January 14, 2008, 13:50 hr, normalized by $B_{rms} = 2.5$ nT, $V_A=75$ km/s. 
Distance $x$ is normalized by $L.$ MVF components are identified as L (\textcolor{red}{red}, solid), 
M( \textcolor{ForestGreen}{green}, dotted), N (\textcolor{blue}{blue}, dashed).}
\label{fig1}
\end{figure}

Increments $\delta \bB(\br),\delta \bu(\br)$ are considered to be ``large'' for our 
criterion when their 
magnitudes both exceed 1.5 of their rms values. Using this threshold, we identify possible reconnection events 
in both datasets. See complete catalogues in \cite{Supplement}. Many of the candidate events in both datasets 
resemble the ``double-step'' magnetic reversals bounded by near-discontinuities, which Gosling tends to select 
with his original criteria \cite{Gosling12}. However, we also see events with more gradual reversals over inertial-range 
scales in both datasets.  In Fig.~\ref{fig1} we show  \\ 

\noindent 
events of this latter type. The vectors have been rotated into the 
minimum-variance-frame (MVF) of the magnetic field \cite{SonnerupCahill67}, calculated over the reversal 
region. The 
velocities 
here (and in all following plots) 
are in a 
frame moving with the local mean plasma velocity. Both events are inertial-range scale,  occupying 
an interval of length 0.1 in the MHD simulation and 2-3 min in the solar wind case. Although they do not 
have a ``double-step'' magnetic structure, 
these two events do show the 
features characteristic of magnetic reconnection. There appears to be a reconnecting field component 
and an associated Alfv\'enic outflow jet in the  $L$-direction of maximum variance. A weak inflow is seen in the 
$N$-direction of minimum variance, which is usually interpreted as across the reconnection ``current sheet.'' 
The $M$-direction of intermediate variance is nominally the guide-field direction, which in both events appears 
rather weak and variable. 

The MHD event shown in Fig.~1, top panels, arises from passage of the sampled 1D cut close to a large, helical magnetic 
flux-rope appearing in the simulation. The maximum field strength in the rope is 8 times the rms strength in the 
database. Plotted in Fig.~3 is the original 1D spatial cut, the magnetic cloud, and nominally incoming and outgoing
field-lines along the N- and L-directions of the MVF.  There is a clear magnetic reversal, with incoming lines in 
the flux-rope twisting clockwise and into the page, but incoming lines to the left pointing out of the page. 
The field-line geometry is, however, quite complex since the lines exhibit the
 stochastic wandering assumed 
 in the Lazarian-Vishniac 
  \\

\begin{figure}[h!]
\includegraphics[width=0.8\columnwidth]{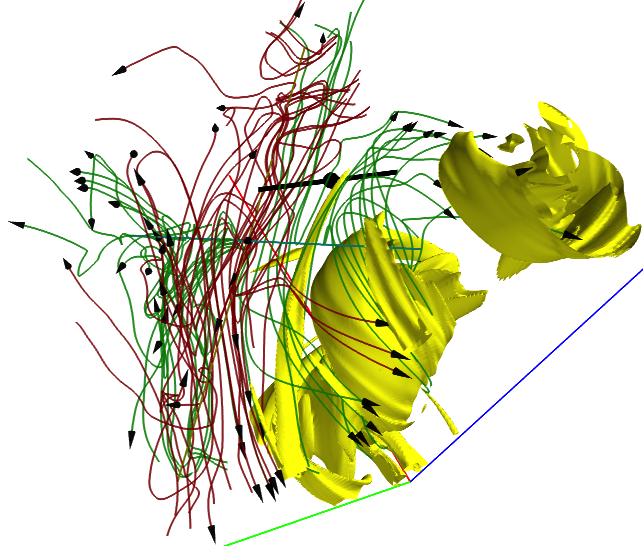}
\caption{B-isosurface at half-maximum value 1.11 in \textcolor{GreenYellow}{yellow}.  
$\bf{\bB}$-lines sampled along N-direction in \textcolor{ForestGreen}{green} and L-directions in \textcolor{BrickRed}{red}.
The original 1D spatial cut is the thick {\bf black} line.} 
\label{fig2}
\end{figure}

\noindent  theory of turbulent reconnection \cite{LazarianVishniac99}. 
Fig.~2 and all other spatial plots in this letter are available as 3D PDF's \cite{Supplement}.  

To identify large-scale geometry, it is necessary to spatially coarse-grain (low-pass filter) the magnetic field. 
For the theoretical basis of this coarse-graining approach to turbulent reconnection, see \cite{EyinkAluie06,Eyink15}.
Here we apply a box-filter with half-width $L$ to obtain coarse-grained fields $\bar{\bB}$, $\bar{\bu},$ from which all 
inertial- and dissipation-range eddies are eliminated. The nature of the database event as large-scale reconnection becomes 
more evident in Fig.~3, which plots the lines of $\bar{\bB}.$ A central ``X-point'' at (2.84, 1.31, 5.73) was located by eye and a 
new MVF calculated in a sphere of radius $L$ around that point. (This frame is rotated by $\sim 20^\circ$ in all three 
directions relative to the MVF for the original 1D cut, but furthermore the $M$- and $N$-directions are exchanged). 
Field-lines are plotted at regular intervals along the $L$- and $N$-axes through the point. The plasma flow is incoming 
along the $N$-direction and outgoing along the $L$-direction,
and the magnetic structure is clearly X-type,  with length $\sim$ 0.4-0.6 ($L$-direction) and width 
$\sim$ 0.15-0.2 ($N$-direction). Reconnection events observed in the solar wind also appear to be X-type 
\cite{Gosling12}, although this structure has generally been interpreted in terms of Petschek reconnection.  

\begin{figure}[h!]
\includegraphics[width=0.8\columnwidth]{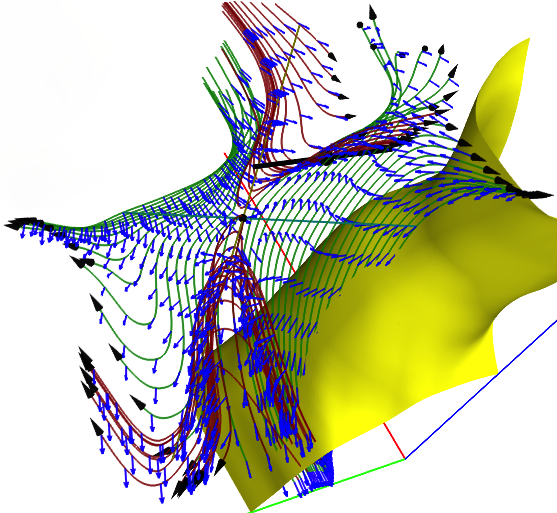}
\caption{Same as Fig.~2, except for $\bar{\bB}$ rather than $\bB.$ The isosurface is for the half-maximum 
value $|\bar{\bB}|=0.301$. The \textcolor{blue}{{\bf blue}} vectors on the 
field-lines are $\bar{\bu}$ (in the local plasma frame).} 
\label{fig3}
\end{figure}

\begin{figure}[h!]
\includegraphics[width=0.63\columnwidth,height=0.54\columnwidth]{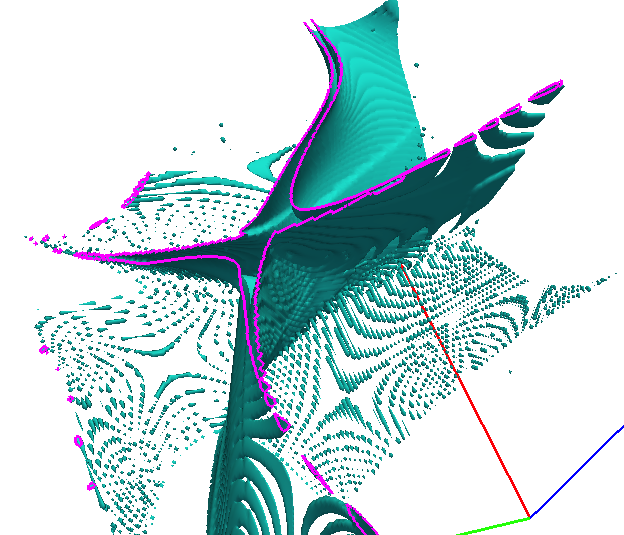}
\caption{The quasi-separatrix layer $Q_\perp=32$ in \textcolor{BlueGreen}{cyan}, and its cross-section in a plane 
normal to the $M$-direction in \textcolor{ElectricPink}{magenta}.} 
\label{fig4}
\end{figure}

In fact, the field-line geometry of $\bar{\bB}$ in the MHD event is more complex than a single ``X-point.'' The 
complete structure is revealed by calculating the ``perpendicular squashing factor'' $Q_\perp,$ a quantity  
devised to identify field-lines with rapidly changing connectivity in the solar photosphere and corona \cite{Titov07}. 
We consider the $Q_\perp$-factor for the field-lines of $\bar{\bB}$ which begin and end on a sphere of radius 1.6$L$ around 
the nominal ``X-point'' in Fig.~\ref{fig3}.  The $Q_\perp$-isosurface in Fig.~\ref{fig4} reveals 
a ``quasi-separatrix layer'' (QSL) whose cross-section has a clear $X$-type structure. The ``hyperbolic 
flux-tube'' (HFT) extending along the centers of these $X$'s has length about 0.54 and is aligned 
approximately with the $M$-direction, to within about $35^\circ.$ An HFT is the modern version of an ``X-line'' 
for 3D reconnection, which does not usually admit true separatrices and X-lines, and an HFT has the 
same observational consequences as an X-line. It is thus interesting that very large-scale reconnection events 
in the solar wind (above integral scales) appear to have very extended $X$-lines, based on observations by multiple spacecraft 
\cite{Phanetal09}. 

Careful examination of the dynamics of this MHD event verifies that it is indeed magnetic reconnection,
and fundamentally influenced by turbulence. The magnetic flux-rope and the associated QSL persist over 
the entire time (0 to 2.56) of the database, drifting slowly with the plasma. The QSL and MVF also 
slowly rotate in time, with the MVF directions rotated through total angles $\sim 40^\circ$ at the final time and also 
the $M$- and $N$-directions exchanged around time $2.0.$ The time required for a plasma fluid element 
in the reconnection region to be carried out by the exhausts with velocities $\sim 0.3$-$0.4$ also happens 
to be about 2.0. Despite the high conductivity of the simulation, standard flux-freezing is violated in this event 
due to the turbulent phenomenon of ``spontaneous stochasticity'', as we now verify. The exact stochastic 
flux-freezing theorem for resistive MHD \cite{Eyink09} (which generalizes ordinary flux-freezing), states that 
field-lines of the fine-grained magnetic field $\bB$ are ``frozen-in'' to the stochastic trajectories solving the 
Langevin equation  
\begin{equation} d\bx/dt =\bu(\bx,t)+\sqrt{2\lambda} \ d\boeta(t), \label{eq1} \end{equation} 
where $\lambda= \eta c^2/4\pi$ is magnetic diffusivity and $\boeta(t)$ is a 3D Gaussian white-noise. The 
many ``virtual'' field-vectors $\tilde{\bB}$ which arrive to the same final point must be averaged to obtain the
physical magnetic field $\bB$ at that point. We have chosen a point $\bx_f$ in the outflow jet in the $+L$-dir-

\begin{figure}[h!]
\includegraphics[width=0.8\columnwidth,height=0.85\columnwidth]{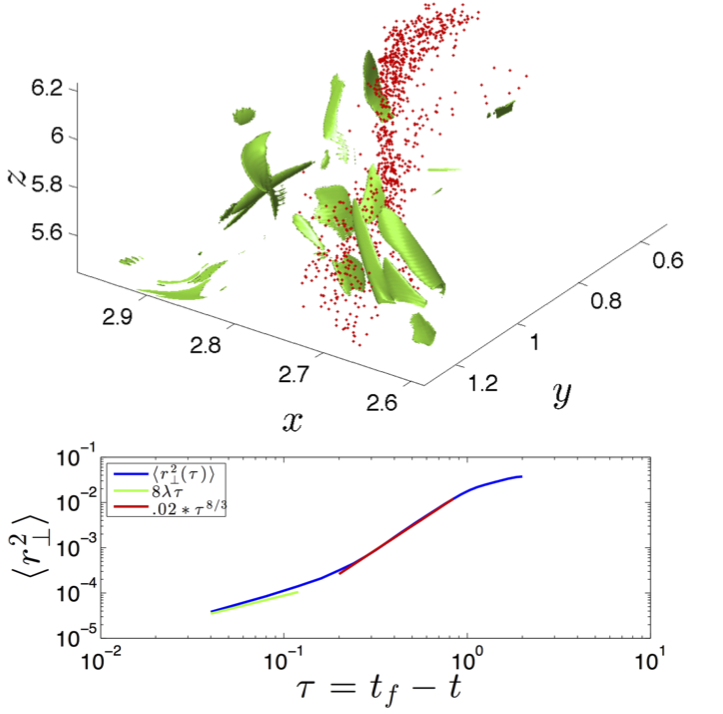}
\caption{{\it Top:} Vorticity isosurfaces at the half-maximum value $\omega=69.1$ in \textcolor{green}{green}. In \textcolor{red}{red}, origin points 
at time $t=0$ of magnetic field at final point $(\bx_f,t_f)=(3.31, 0.083, 6.07,2.00)$. {\it Bottom:} Backward mean-square dispersion 
$\langle r_\perp^2(\tau)\rangle$ orthogonal to $L$-direction as \textcolor{blue}{blue} line. Reference curve $\langle r_\perp^2\rangle=8\lambda\tau$
in \textcolor{green}{green}, and $\langle r_\perp^2\rangle= 0.02\tau^{8/3}$ in \textcolor{red}{red}.} 
\label{fig5}
\end{figure}

\noindent
ection at time $t_f=2$ and solved (\ref{eq1}) backward in time to $t_0=0,$ to find the positions of the initial points 
whose magnetic field vectors arrive at $(\bx_f,t_f).$ This ensemble of points, plotted in Fig.~5, 
is widely dispersed in space. This disagrees with the predictions of standard flux-freezing, which implies that
the ensemble should be close to a single point. In the lower panel of Fig.~5 is plotted the mean-square 
dispersion of this ensemble perpendicular to the L-direction, $\langle r_\perp^2\rangle$ as a function of reversed time
$\tau=t_f-t.$ Consistent with previous results \cite{Eyinketal13}, the (backward) growth of perpendicular 
dispersion is diffusive $\langle r_\perp^2(\tau)\rangle\sim 8\lambda \tau $ for very small $\tau$ but then becomes super-ballistic, due to 
turbulent Richardson dispersion. As argued in \cite{Eyinketal11}, the 
perpendicular spread in the time to exit with the outflow, $\sqrt{\langle r_\perp^2(2)\rangle}\sim 0.19,$ is close to the width of the 
reconnection region. This zone has both the width and the turbulent structure proposed in \cite{LazarianVishniac99},
as can be seen also in Fig.~\ref{fig5} which plots in green the isosurfaces of the fine-grained current magnitude
at half-maximum. There is a spatial distribution of many current sheets rather than a single large current sheet, 
as in laminar reconnection, and none of the sheets is located precisely at the QSL shown in Fig.~\ref{fig4}.  
See \cite{Supplement} for 3D plots.

The breakdown of standard flux-freezing is one evidence of reconnection in this event \cite{Greene93}. 
We have also verified that there is topology change of the lines of both fine-grained and coarse-grained magnetic 
fields. To show this, we decorate initial field-lines of either $\bB$ or $\bar{\bB}$ at time $t=0$ with 
a sequence of plasma fluid elements and then follow each element moving with the local velocity $\bu$ forward 
in time to $t=2.$ We find that the plasma elements which initially resided on the same line at $t=0$ 
end up on distinct lines at time $t=2$ and some of these lines are outgoing in the $+L$-direction and 
others in the $-L$-direction. For movies, see \cite{Supplement}. We have also determined the 
average reconnecting electric field $E_{{\rm rec}}$ for the 
large-scale magnetic field $\bar{\bB},$ using a voltage measure proposed in \cite{Kowaletal09}.
We find that $E_{{\rm rec}}\sim 0.01 v_A B$ in terms of local upstream values
$v_A$ and $B$. Furthermore, at the length scale $L$ of $\bar{\bB},$
most of $E_{{\rm rec}}$ is supplied by turbulence-induced electric fields and resistivity 
gives only a tiny contribution, always more than an order of magnitude smaller.  These 
and many other detailed results for this event will be presented  elsewhere.
One finding is that this inertial-range event is not only highly 3D but also non-stationary in time. 
While the outflow jets are quite stable over time, the inflow is ``gusty'', with variable magnitude and direction 
veering in the $N$-$M$-plane (so that it is often in the nominal guide-field or $M$-direction). It is 
thus difficult to define an operationally meaningful ``reconnection speed''.

The main purpose of this Letter has been to present an example of inertial-range reconnection in MHD turbulence, 
to clarify its observational signatures. While a fluid description is surely applicable only to scales much larger than plasma 
micro-scales (e.g. the ion gyroradius in the collisionless solar wind), our simulation is remarkably successful in reproducing 
observed features of large-scale solar-wind reconnection, together with crucial turbulent effects supporting theoretical predictions in \cite{LazarianVishniac99,Eyinketal11}. 
The characteristics are expected to change with length-scale, e.g. reconnection deeper within the inertial-range should have stronger guide-fields/smaller 
magnetic shear-angles \cite{LazarianVishniac99,Goslingetal07}. We also note some differences between the current MHD database 
and the solar wind, as our simulation is incompressible and isothermal, whereas the solar wind is slightly compressible and 
reconnection events there (including that in Fig. 1) often show enhancements of proton density and temperature in the reconnection zone. 
Furthermore, our MHD simulation has no mean magnetic field and is close to balance between Alfv\'en waves 
propagating parallel and anti-parallel to field-lines, whereas the solar wind has a moderate mean field and the high-speed stream studied 
in this work is dominated by Alfv\'en waves propagating outward from the sun. The influence of these differences should be explored 
in future work seeking to explain large-scale solar wind reconnection in detail within an MHD turbulence framework.


	\subsection{Acknowledgments}
		This work was supported by  NSF grants CDI-II: CMMI 0941530 and AST 1212096.

	\bibliography{inertial-reconnect}

\end{document}